\documentclass{article}
\usepackage{amssymb}
\begin{document}
\title{Constants of motion associated with alternative Hamiltonians}

\author{G.F.\ Torres del Castillo \\ Departamento de F\'isica Matem\'atica, Instituto de Ciencias \\
Universidad Aut\'onoma de Puebla, 72570 Puebla, Pue., M\'exico}

\maketitle

\begin{abstract}
It is shown that if a non-autonomous system of $2n$ first-order ordinary differential equations is expressed in the form of the Hamilton equations in terms of two different sets of coordinates, $(q_{i}, p_{i})$ and $(Q_{i}, P_{i})$, then the determinant and the trace of any power of a certain matrix formed by the Poisson brackets of the $Q_{i}, P_{i}$ with respect to $q_{i}, p_{i}$, are constants of motion.
\end{abstract}

\noindent PACS numbers: 45.20.Jj, 02.30.Hq

\section{Introduction}
Given a second-order ordinary differential equation (ODE) (which may be the equation of motion of a mechanical system with one degree of freedom), there exists an infinite number of Lagrangians that reproduce the given equation. If $L(q, \dot{q}, t)$ and $L'(q, \dot{q}, t)$ are two of such Lagrangians, then $\partial^{2} L'/\partial \dot{q}^{2}$ divided by $\partial^{2} L/\partial \dot{q}^{2}$ is a constant of motion (see, e.g., Refs.\ \cite{Wi,CR}, and the references cited therein). Conversely, given a Lagrangian and a constant of motion, one can combine them to find a Lagrangian alternative to the one already known.

In a similar way, if the Lagrangians, $L(q_{i}, \dot{q}_{i}, t)$ and $L'(q_{i}, \dot{q}_{i}, t)$, lead to two equivalent systems of $n$ second-order ODEs (with $n \geqslant 2$), then the trace of any power of the product of the $n \times n$ matrix $(\partial^{2} L'/\partial \dot{q_{i}} \partial \dot{q}_{j})$ by the inverse of $(\partial^{2} L/\partial \dot{q_{i}} \partial \dot{q}_{j})$ is a constant of motion \cite{HH}. An important difference between the cases $n = 1$ and $n > 1$, is that there exist systems of two or more second-order ODEs that cannot be obtained from a Lagrangian (see, e.g., Ref.\ \cite{AT}).

On the other hand, regarding the Hamiltonian formalism, any system of $2n$ autonomous (i.e., time-independent) first-order EDOs can be expressed in Hamiltonian form, in an infinite number of different ways (see, e.g., Ref.\ \cite{GE} an the references cited therein). That is, given a system of $2n$ first-order ODEs,
\begin{equation}
\dot{y}^{\alpha} = f^{\alpha}(y^{1}, \ldots, y^{2n}), \qquad \alpha = 1,2, \ldots, 2n, \label{sys1}
\end{equation}
there exists an infinite number of pairs formed by a nonsingular matrix, $(\sigma^{\alpha \beta})$, and a real-valued function, $H$, such that
\begin{equation}
\dot{y}^{\alpha} = \sigma^{\alpha \beta} \frac{\partial H}{\partial y^{\beta}} \label{1}
\end{equation}
(here and henceforth, there is summation over repeated indices) and
\begin{equation}
\sigma^{\mu \gamma} \frac{\partial \sigma^{\alpha \beta}}{\partial y^{\mu}} + \sigma^{\mu \alpha} \frac{\partial \sigma^{\beta \gamma}}{\partial y^{\mu}} + \sigma^{\mu \beta} \frac{\partial \sigma^{\gamma \alpha}}{\partial y^{\mu}} = 0. \label{symp}
\end{equation}
(This last condition allows one to define a Poisson bracket that satisfies the Jacobi identity.) If the system (\ref{sys1}) is also expressed in the form
\begin{equation}
\dot{y}^{\alpha} = \sigma'{}^{\alpha \beta} \frac{\partial H'}{\partial y^{\beta}}, \label{2}
\end{equation}
then the determinant and the trace of any power of the matrix $S = (S^{\alpha}_{\beta})$, defined by
\begin{equation}
S^{\alpha}_{\beta} = \sigma^{\alpha \gamma} \sigma'{}_{\gamma \beta}, \label{das}
\end{equation}
where $(\sigma'{}_{\alpha \beta})$ is the inverse of $(\sigma'{}^{\alpha \beta})$, is a constant of motion (see Ref.\ \cite{Da} and the references cited therein).

In the case of a {\em non-autonomous}\/ system of two first-order ODEs,
\[
\dot{y}^{1} = f^{1}(y^{1}, y^{2}, t), \qquad \dot{y}^{2} = f^{2}(y^{1}, y^{2}, t),
\]
there exists an infinite number of coordinate systems, $q, p$, such that these equations can be written in the canonical form
\begin{equation}
\dot{q} = \frac{\partial H}{\partial p}, \qquad \dot{p} = - \frac{\partial H}{\partial q}, \label{can1}
\end{equation}
for some function $H$. If $Q, P$ is another coordinate system such that
\begin{equation}
\dot{Q} = \frac{\partial H'}{\partial P}, \qquad \dot{P} = - \frac{\partial H'}{\partial Q}, \label{can2}
\end{equation}
for some alternative Hamiltonian $H'$, then $\{ Q, P \}$ is a constant of motion, where $\{ \; , \; \}$ is the Poisson bracket defined by the coordinates $q, p$ \cite{Ge}. Note that, by contrast with the previous cases, in Eqs.\ (\ref{can1}) and (\ref{can2}) there are two coordinate systems involved.

In this paper we prove, in an elementary manner, that if a non-autonomous system of $2n$ first-order ODEs is expressed in the form of the Hamilton equations in terms of two different sets of coordinates, $(q_{i}, p_{i})$ and $(Q_{i}, P_{i})$, then the determinant and the trace of any power of a certain matrix formed by the Poisson brackets of the $Q_{i}, P_{i}$ with respect to $q_{i}, p_{i}$, are constants of motion. We also show that this result contains all those mentioned above.

In Section 2 we give a simple derivation of the result presented in Ref.\ \cite{HH}, related to the Lagrangian formalism. In Section 3 we consider the general case of a non-autonomous system of $2n$ first-order ODEs, showing that if the system can be expressed in Hamiltonian form in terms of different sets of coordinates, not related by canonical transformations, several constants of motion can be obtained, and in Section 4 we show how this result reduces to those previously established.

\section{Equivalent Lagrangians}
As in Ref.\ \cite{HH}, we shall consider a system of $n$ second-order ODEs,
\begin{equation}
\ddot{q}_{i} = F_{i}(q_{j}, \dot{q}_{j}, t), \qquad i = 1, 2, \ldots, n, \label{sys2}
\end{equation}
that can be expressed as the Euler--Lagrange equations for some regular Lagrangian, $L(q_{i}, \dot{q}_{i}, t)$. That is, we assume that Eqs.\ (\ref{sys2}) are equivalent to the Euler--Lagrange equations
\begin{equation}
\frac{\partial^{2} L}{\partial \dot{q}_{j} \partial \dot{q}_{i}} \ddot{q}_{j} + \frac{\partial^{2} L}{\partial q_{j} \partial \dot{q}_{i}} \dot{q}_{j} + \frac{\partial^{2} L}{\partial t \partial \dot{q}_{i}} - \frac{\partial L}{\partial q_{i}} = 0, \label{el}
\end{equation}
$i = 1, 2, \ldots, n$. As usual, the regularity of $L$ means that $\det (M_{ij}) \not= 0$, with
\begin{equation}
M_{ij} \equiv \frac{\partial^{2} L}{\partial \dot{q}_{i} \partial \dot{q}_{j}}, \label{reg}
\end{equation}
and we will assume that all the partial derivatives commute, so that $(M_{ij})$ is a symmetric $n \times n$ matrix.

Taking the partial derivative of Eqs.\ (\ref{el}) with respect to $\dot{q}_{k}$, we find that the $n^{2}$ equations
\begin{equation}
\frac{{\rm d} M_{ik}}{{\rm d} t} + M_{ij} \frac{\partial F_{j}}{\partial \dot{q}_{k}} + \frac{\partial^{2} L}{\partial q_{k} \partial \dot{q}_{i}} - \frac{\partial^{2} L}{\partial q_{i} \partial \dot{q}_{k}} = 0 \label{cond1}
\end{equation}
must hold as a consequence of Eqs.\ (\ref{sys2}). For $n \geqslant 2$, we can decompose this system of equations into a symmetric and an antisymmetric part, viz.,
\begin{equation}
\frac{{\rm d} M_{ik}}{{\rm d} t} + \frac{1}{2} \left( M_{ij} \frac{\partial F_{j}}{\partial \dot{q}_{k}} + M_{kj} \frac{\partial F_{j}}{\partial \dot{q}_{i}}\right) = 0 \label{sym}
\end{equation}
and
\begin{equation}
\frac{1}{2} \left( M_{ij} \frac{\partial F_{j}}{\partial \dot{q}_{k}} - M_{kj} \frac{\partial F_{j}}{\partial \dot{q}_{i}}\right) + \frac{\partial^{2} L}{\partial q_{k} \partial \dot{q}_{i}} - \frac{\partial^{2} L}{\partial q_{i} \partial \dot{q}_{k}} = 0. \label{ant}
\end{equation}
(When $n = 1$, Eqs.\ (\ref{ant}) reduce to the identity $0 = 0$.)

It is convenient to define the functions
\[
\Phi_{ij} \equiv \frac{1}{2} \frac{\partial F_{j}}{\partial \dot{q}_{i}},
\]
and the $n \times n$ matrices $M \equiv (M_{ij})$ and $\Phi \equiv (\Phi_{ij})$, so that Eqs.\ (\ref{sym}) are equivalent to the matrix equation
\begin{equation}
\frac{{\rm d} M}{{\rm d} t} = - (M \Phi^{{\rm t}} + \Phi M), \label{defi}
\end{equation}
where $\Phi^{{\rm t}}$ denotes the transpose of $\Phi$. Hence,
\[
\frac{{\rm d} M^{-1}}{{\rm d} t} = - M^{-1} \frac{{\rm d} M}{{\rm d} t} M^{-1} = \Phi^{{\rm t}} M^{-1} + M^{-1} \Phi.
\]

If the Lagrangian $L'(q_{i}, \dot{q}_{i}, t)$ also leads to Eqs.\ (\ref{sys2}), a relation analogous to Eq.\ (\ref{defi}), {\em with the same}\/ $\Phi$, must also hold for the matrix $M'$ with entries $M'{}_{ij} \equiv \partial^{2} L'/\partial \dot{q_{i}} \partial \dot{q}_{j}$; thus, letting
\begin{equation}
\Lambda \equiv M' M^{-1}, \label{lambda}
\end{equation}
 we obtain
\begin{equation}
\frac{{\rm d} \Lambda}{{\rm d} t} = M' (\Phi^{{\rm t}} M^{-1} + M^{-1} \Phi) - (M' \Phi^{{\rm t}} + \Phi M') M^{-1} = \Lambda \Phi - \Phi \Lambda. \label{lax}
\end{equation}
From this last equation one readily finds that, for any integer $N$ (including negative values),
\[
\frac{{\rm d} \Lambda^{N}}{{\rm d} t} = \Lambda^{N} \Phi - \Phi \Lambda^{N},
\]
which implies that
\begin{equation}
\frac{{\rm d} ({\rm tr}\, \Lambda^{N})}{{\rm d} t} = 0. \label{consl}
\end{equation}
That is, the trace of $\Lambda^{N}$ is a constant of motion (though it may be a trivial constant) (cf.\ the proofs given in Refs.\ \cite{HH,FN} and the references cited therein). Note that the number of functionally independent traces of powers of $\Lambda$ cannot exceed $n$ (for instance, if $\Lambda$ is diagonalizable, the trace of $\Lambda^{N}$ is equal to the sum $\lambda_{1}{}^{N} + \lambda_{2}{}^{N} + \cdots + \lambda_{n}{}^{N}$, where the $\lambda_{s}$ are the eigenvalues of $\Lambda$).

Making use of the formula
\begin{equation}
\frac{{\rm d} \ln |\det A|}{{\rm d} t} = {\rm tr}\, \left( A^{-1} \frac{{\rm d} A}{{\rm d} t} \right) \label{det}
\end{equation}
(which can be derived, e.g.,  from the well-known relation $\det \exp B = \exp {\rm tr}\, B$), from Eq.\ (\ref{lax}) we find that also $\det \Lambda$ is a constant of motion.

\section{Equivalent Hamiltonians}
Since with each regular Lagrangian, $L(q_{i}, \dot{q}_{i}, t)$, there is an associated Hamiltonian, $H(q_{i}, p_{i}, t)$, that leads to a system of $2n$ first-order ODEs equivalent to the Euler--Lagrange equations (\ref{el}), two alternative Lagrangians, $L$ and $L'$, corresponding to Eqs.\ (\ref{sys2}), define two alternative Hamiltonians, $H$ and $H'$, which, substituted into the Hamilton equations, will produce two equivalent systems of first-order equations. However, even if we express $L$ and $L'$ in terms of the same coordinates $q_{i}$ (as we did in the preceding section), the Hamilton equations for $H$ and $H'$ involve different conjugate momenta, $p_{i} \equiv \partial L/\partial \dot{q}_{i}$ and $p'{}_{i} \equiv \partial L'/\partial \dot{q}_{i}$; furthermore, as we shall see, the coordinates $(q_{i}, p_{i})$ and $(q_{i}, p'{}_{i})$ {\em need not}\/ be related by means of a canonical transformation. Hence, in order to compare the results of Sec.\ 2 with those obtained by means of the Hamiltonian formalism, in what follows it will be convenient to consider two coordinate systems, $(q_{i}, p_{i})$ and $(Q_{i}, P_{i})$, not necessarily related by a canonical transformation.

Assuming that the Hamilton equations
\begin{equation}
\dot{q_{i}} = \frac{\partial H}{\partial p_{i}}, \qquad \dot{p_{i}} = - \frac{\partial H}{\partial q_{i}}, \label{ham1}
\end{equation}
are equivalent to
\begin{equation}
\dot{Q_{i}} = \frac{\partial H'}{\partial P_{i}}, \qquad \dot{P_{i}} = - \frac{\partial H'}{\partial Q_{i}},
\label{ham2}
\end{equation}
for some Hamiltonian functions $H$ and $H'$, guided by the results mentioned in the Introduction, we shall consider the $2n \times 2n$ matrix $S = (S^{\alpha}_{\beta})$, defined by
\begin{equation}
S^{\alpha}_{\beta} \equiv \{ y^{\alpha}, y^{\gamma} \} \, \epsilon_{\gamma \beta} \label{ese}
\end{equation}
($\alpha, \beta, \gamma = 1, 2, \ldots, 2n$), where
\begin{equation}
(y^{1}, \ldots, y^{n}, y^{n+1}, \ldots, y^{2n}) \equiv (Q_{1}, \ldots, Q_{n}, P_{1}, \ldots, P_{n}), \label{QP}
\end{equation}
$\{ \; , \; \}$ denotes the Poisson bracket defined by the coordinates $(q_{i}, p_{i})$,
\begin{equation}
\{ f, g \} = \frac{\partial f}{\partial q_{i}} \frac{\partial g}{\partial p_{i}} - \frac{\partial f}{\partial p_{i}} \frac{\partial g}{\partial q_{i}}, \label{pbe}
\end{equation}
and $(\epsilon_{\alpha \beta})$ is the block matrix
\begin{equation}
(\epsilon_{\alpha \beta}) \equiv \left( \begin{array}{cc} 0 & - I \\ I & 0 \end{array} \right), \label{can}
\end{equation}
where $I$ is the $n \times n$ unit matrix.

Letting
\begin{equation}
(x^{1}, \ldots, x^{n}, x^{n+1}, \ldots, x^{2n}) \equiv (q_{1}, \ldots, q_{n}, p_{1}, \ldots, p_{n}) \label{qp}
\end{equation}
[cf.\ Eq.\ (\ref{QP})], the Poisson bracket (\ref{pbe}) is expressed as
\begin{equation}
\{ f, g \} = \epsilon^{\alpha \beta} \frac{\partial f}{\partial x^{\alpha}} \frac{\partial g}{\partial x^{\beta}}, \label{pb}
\end{equation}
where $(\epsilon^{\alpha \beta})$ is the inverse of the matrix $(\epsilon_{\alpha \beta})$, i.e.,
\begin{equation}
(\epsilon^{\alpha \beta}) \equiv \left( \begin{array}{cc} 0 & I \\ - I & 0 \end{array} \right).
\end{equation}
The main result of this paper can be expressed as follows.

\noindent {\bf Proposition.} The matrix $S$ defined in (\ref{ese}) satisfies the equation
\begin{equation}
\frac{{\rm d} S}{{\rm d} t} = US - SU, \label{laxh}
\end{equation}
where $U = (U^{\alpha}_{\beta})$ is the $2n \times 2n$ matrix defined by
\[
U^{\alpha}_{\beta} \equiv \epsilon^{\alpha \gamma} \frac{\partial^{2} H'}{\partial y^{\gamma} \partial y^{\beta}}.
\]

\noindent {\em Proof.} As is well known, from the definition of the Poisson bracket and the Jacobi identity it follows that
\[
\frac{{\rm d} \{ f, g \}}{{\rm d} t} =  \left\{ f, \frac{{\rm d} g}{{\rm d} t} \right\} + \left\{ \frac{{\rm d} f}{{\rm d} t}, g \right\}
\]
(which is essentially the Poisson theorem about constants of motion), hence, from the definition (\ref{ese}) we have
\[
\frac{{\rm d} S^{\alpha}_{\beta}}{{\rm d} t} = \{ y^{\alpha}, \dot{y}^{\gamma} \} \, \epsilon_{\gamma \beta} + \{ \dot{y}^{\alpha}, y^{\gamma} \} \, \epsilon_{\gamma \beta}.
\]
With the aid of the notation (\ref{QP}), the Hamilton equations (\ref{ham2}) can be written as
\begin{equation}
\dot{y}^{\alpha} = \epsilon^{\alpha \beta} \frac{\partial H'}{\partial y^{\beta}}, \label{adi}
\end{equation}
therefore, making use of (\ref{pb}) and the chain rule,
\begin{eqnarray*}
\frac{{\rm d} S^{\alpha}_{\beta}}{{\rm d} t} & = & \left\{ y^{\alpha}, \epsilon^{\gamma \mu} \frac{\partial H'}{\partial y^{\mu}} \right\} \epsilon_{\gamma \beta} + \left\{ \epsilon^{\alpha \mu} \frac{\partial H'}{\partial y^{\mu}}, y^{\gamma} \right\} \epsilon_{\gamma \beta} \\
& = & \epsilon^{\gamma \mu} \epsilon_{\gamma \beta} \epsilon^{\rho \sigma} \frac{\partial y^{\alpha}}{\partial x^{\rho}} \left( \frac{\partial}{\partial x^{\sigma}} \frac{\partial H'}{\partial y^{\mu}} \right) + \epsilon^{\alpha \mu} \epsilon_{\gamma \beta} \epsilon^{\rho \sigma} \left( \frac{\partial}{\partial x^{\rho}} \frac{\partial H'}{\partial y^{\mu}} \right) \frac{\partial y^{\gamma}}{\partial x^{\sigma}} \\
& = & - \delta^{\mu}_{\beta} \epsilon^{\rho \sigma} \frac{\partial y^{\alpha}}{\partial x^{\rho}} \frac{\partial y^{\lambda}}{\partial x^{\sigma}} \frac{\partial^{2} H'}{\partial y^{\lambda} \partial y^{\mu}} + \epsilon^{\alpha \mu} \epsilon_{\gamma \beta} \epsilon^{\rho \sigma} \frac{\partial y^{\lambda}}{\partial x^{\rho}} \frac{\partial y^{\gamma}}{\partial x^{\sigma}} \frac{\partial^{2} H'}{\partial y^{\lambda} \partial y^{\mu}} \\
& = & - \{ y^{\alpha}, y^{\lambda} \} \frac{\partial^{2} H'}{\partial y^{\lambda} \partial y^{\beta}} + \epsilon^{\alpha \mu} \epsilon_{\gamma \beta} \{ y^{\lambda}, y^{\gamma} \} \frac{\partial^{2} H'}{\partial y^{\lambda} \partial y^{\mu}} \\
& = & - S^{\alpha}_{\mu} \epsilon^{\mu \lambda} \frac{\partial^{2} H'}{\partial y^{\lambda} \partial y^{\beta}} + \epsilon^{\alpha \mu} S^{\lambda}_{\beta} \frac{\partial^{2} H'}{\partial y^{\lambda} \partial y^{\mu}} \\
& = & - S^{\alpha}_{\mu} U^{\mu}_{\beta} + U^{\alpha}_{\lambda} S^{\lambda}_{\beta},
\end{eqnarray*}
thus proving the validity of (\ref{laxh}).

As in the case of Eq.\ (\ref{lax}), from Eq.\ (\ref{laxh}) it follows that, for $N = \pm 1, \pm 2, \ldots \,$, \begin{equation}
\frac{{\rm d} ({\rm tr}\, S^{N})}{{\rm d} t} = 0 \label{consh}
\end{equation}
and that $\det S$ is also a constant of motion.

From Eqs.\ (\ref{ese}), (\ref{QP}), and (\ref{can}) one finds that $S$ is the block matrix
\begin{equation}
S = \left( \begin{array}{cc} (\{ Q_{i}, P_{j} \}) & - (\{ Q_{i}, Q_{j} \}) \\[1ex] (\{ P_{i}, P_{j} \}) & - (\{ P_{i}, Q_{j} \}) \end{array} \right). \label{eseblo}
\end{equation}
Note that $S$ is the unit matrix if and only if the coordinates $Q_{i}, P_{i}$ are related to $q_{i}, p_{i}$ by means of a canonical transformation.

\section{Connection with previous results}
In the case where $n = 1$, the matrix (\ref{eseblo}) reduces to the $2 \times 2$ matrix
\[
S = \left( \begin{array}{cc} \{ Q, P \} & 0 \\[1ex] 0 & \{ Q, P \} \end{array} \right),
\]
which is proportional to the unit matrix; therefore, the right-hand side of Eq.\ (\ref{laxh}) is always equal to zero. Hence, $\{ Q, P \}$ is a constant of motion, and all the traces ${\rm tr}\, S^{N}$, as well as $\det S$, are functions of this constant. Note that, in this case, the only conditions on the coordinates $q, p$ and $Q, P$ are Eqs.\ (\ref{ham1}) and (\ref{ham2}); moreover, we do not have to assume that these equations come from some Lagrangians.

Now we shall show explicitly that if we have two equivalent Lagrangians, $L$ and $L'$, the constants of motion (\ref{consl}) are, up to a constant factor, those obtained from Eqs.\ (\ref{consh}), considering the Hamiltonians corresponding to $L$ and $L'$.

In fact, starting from the Lagrangian $L(q_{i}, \dot{q}_{i}, t)$, the standard expression
\begin{equation}
p_{i} = \frac{\partial L}{\partial \dot{q}_{i}} \label{mom}
\end{equation}
gives $p_{i}$ as a function of $q_{i}, \dot{q}_{i}$, and $t$; hence, making use of the definition (\ref{reg}),
\begin{eqnarray}
{\rm d} p_{i} & = & \frac{\partial^{2} L}{\partial q_{j} \partial \dot{q}_{i}} {\rm d} q_{j} + \frac{\partial^{2} L}{\partial \dot{q}_{j} \partial \dot{q}_{i}} {\rm d} \dot{q}_{j} + \frac{\partial^{2} L}{\partial t \partial \dot{q}_{i}} {\rm d} t \nonumber \\
& = & M_{ij} {\rm d} \dot{q}_{j} + \frac{\partial^{2} L}{\partial q_{j} \partial \dot{q}_{i}} {\rm d} q_{j} + \frac{\partial^{2} L}{\partial t \partial \dot{q}_{i}} {\rm d} t. \label{dp}
\end{eqnarray}
With an alternative Lagrangian, $L'(q_{i}, \dot{q}_{i}, t)$, we obtain the analogous relation
\begin{equation}
{\rm d} p'{}_{i} = M'{}_{ij} {\rm d} \dot{q}_{j} + \frac{\partial^{2} L'}{\partial q_{j} \partial \dot{q}_{i}} {\rm d} q_{j} + \frac{\partial^{2} L'}{\partial t \partial \dot{q}_{i}} {\rm d} t \label{dpp}
\end{equation}
and, assuming that $L$ is regular, from Eq.\ (\ref{dp}) we find an expression for ${\rm d} \dot{q}_{j}$, which substituted into Eq.\ (\ref{dpp}) gives
\[
{\rm d} p'{}_{i} = M'{}_{ij} (M^{-1})_{jk} {\rm d} p_{k} + {\rm terms\ proportional \ to\ } {\rm d} q_{k} {\rm \ or\ }  {\rm d} t.
\]
Hence, considering $p'{}_{i}$ as a function of $q_{k}, p_{k}$, and $t$,
\[
\frac{\partial p'{}_{i}}{\partial p_{k}} = M'{}_{ij} (M^{-1})_{jk} = \Lambda_{ik},
\]
where we have made use of the definition (\ref{lambda}). Thus, with $Q_{i} = q_{i}$, and $P_{i} = p'{}_{i}$, we find that some of the Poisson brackets appearing in (\ref{eseblo}) are given by
\[
\{ Q_{i}, Q_{j} \} = 0, \qquad \{ Q_{i}, P_{j} \} = \frac{\partial p'{}_{j}}{\partial p_{i}} =  \Lambda_{ji},
\]
which implies that in this case the matrix (\ref{eseblo}) has the form
\[
S = \left( \begin{array}{cc} \Lambda^{{\rm t}} & 0 \\[1ex] (\{ P_{i}, P_{j} \}) & \Lambda \end{array} \right)
\]
and, therefore,
\[
{\rm tr}\, S^{N} = 2 \, {\rm tr}\, \Lambda^{N}.
\]

Finally, we shall show that in the case of an autonomous system (\ref{sys1}), the functions (\ref{ese}) reduce to (\ref{das}) if the coordinates $y^{\alpha}$ appearing in Eqs.\ (\ref{1}) and (\ref{2}) are chosen as in (\ref{QP}). Assuming that the relation between the coordinates (\ref{QP}) and (\ref{qp}) does not involve the time explicitly, with the aid of the chain rule and Eqs.\ (\ref{ham1}) and (\ref{pb}), we obtain
\begin{eqnarray*}
\dot{y}^{\alpha} & = & \frac{\partial y^{\alpha}}{\partial x^{\beta}} \dot{x}^{\beta} = \frac{\partial y^{\alpha}}{\partial x^{\beta}} \epsilon^{\beta \gamma} \frac{\partial H}{\partial x^{\gamma}} \\
& = & \frac{\partial y^{\alpha}}{\partial x^{\beta}} \epsilon^{\beta \gamma} \frac{\partial y^{\mu}}{\partial x^{\gamma}} \frac{\partial H}{\partial y^{\mu}} = \{ y^{\alpha}, y^{\mu} \} \frac{\partial H}{\partial y^{\mu}},
\end{eqnarray*}
which, compared with Eq.\ (\ref{1}), shows that $\sigma^{\alpha \beta} = \{ y^{\alpha}, y^{\beta} \}$. On the other hand, comparison of (\ref{2}) with (\ref{adi}) yields $\sigma'{}^{\alpha \beta} = \epsilon^{\alpha \beta}$. Then, substituting these expressions into Eq.\ (\ref{ese}) one obtains Eq.\ (\ref{das}).

\section{Concluding remarks}
As stressed in Ref.\ \cite{Da}, in the case of an autonomous system (\ref{sys1}), the functions (\ref{das}) are, by construction, the components of a tensor field (with respect to the natural basis defined by the arbitrary coordinates $y^{\alpha}$). By contrast, the definition of the functions (\ref{ese}) involves two different coordinate systems, which are not arbitrary.

Among other things, the results presented here allows us to obtain constants of motion from {\em discrete}\/ or continuous transformations that leave invariant a given set of equations of motion, which need not be canonical.

\end{document}